\documentclass[prd,aps,nofootinbib,floatfix,11pt]{revtex4}

\usepackage{amsmath,graphicx,epsfig,amssymb,dsfont,mathtools}
\usepackage{inputenc}
\usepackage[usenames]{color}
\usepackage{ulem} 
\usepackage{bigstrut}
\usepackage{slashed}
\usepackage{multirow}
\usepackage{subfigure}

\allowdisplaybreaks

\begin{document}
	
\title{On the four-quark operator matrix elements for the lifetime of $\Lambda_{b}$}
	
\author{
Zhen-Xing Zhao$^{1}$~\footnote{Email:zhaozx19@imu.edu.cn},
Xiao-Yu Sun,
Fu-Wei Zhang,
Zhi-Peng Xing$^{2}$~\footnote{Email:zpxing@sjtu.edu.cn}
}

\affiliation{
$^{1}$ School of Physical Science and Technology, \\
Inner Mongolia University, Hohhot 010021, China, \\
$^{2}$ Tsung-Dao Lee Institute, Shanghai Jiao Tong University, Shanghai 200240, China
}

\begin{abstract}
Heavy quark expansion can nicely explain the lifetime of $\Lambda_{b}$.
However, there still exist sizable uncertainties from the four-quark
operator matrix elements of $\Lambda_{b}$ in $1/m_{b}^{3}$ corrections, which describe
the spectator effects. In this work, these four-quark operator matrix
elements are investigated using
full QCD sum rules for the first time. At the QCD level, contributions
from up to dimension-6 four-quark operators are considered.
Our method of calculating high-dimensional operator matrix elements
is promising to be used to resolve the $\Omega_{c}$ lifetime puzzle. 
\end{abstract}

\maketitle

\section{Introduction}

In 2018, LHCb measured the lifetime of $\Omega_{c}^{0}$ using $\Omega_{b}^{-}\to\Omega_{c}^{0}\mu^{-}\bar{\nu}_{\mu}X$
decays and obtained \cite{Aaij:2018dso}
\begin{equation}
\tau(\Omega_{c}^{0})=268\pm24\pm10\pm2\ {\rm fs},
\end{equation}
which is roughly four times as large as 
\begin{equation}
\tau(\Omega_{c}^{0})=69\pm12\ {\rm fs}
\end{equation}
in PDG2018 \cite{Tanabashi:2018oca}. In 2021, LHCb confirmed the
previous measurement using $\Omega_{c}^{0}$ produced directly from
proton-proton collision \cite{LHCb:2021vll}, and in 2022, Belle-II
also reported a similar result using $\Omega_{c}^{0}\to\Omega^{-}\pi^{+}$
decays \cite{Belle-II:2022plj}. These recent measurements indicate
that $\Omega_{c}^{0}$ is not the shortest-lived weakly decaying charmed
baryon, which conflicts with our previous understanding \cite{Lenz:2014jha}. 
Theoretical explanation is highly demanded. 

At present, the standard
framework for understanding weakly decaying heavy flavor hadrons is
heavy quark expansion (HQE) \cite{Khoze:1983yp,Bigi:1991ir,Bigi:1992su,Blok:1992hw,Blok:1992he,Neubert:1997gu,Uraltsev:1998bk,Bigi:1995jr}.
Under this framework, some attempts have been made to resolve the
$\Omega_{c}$ lifetime puzzle \cite{Cheng:2018rkz,Gratrex:2022xpm,Cheng:2023jpz}.
However, there is still a lack of more reliable
calculation based on QCD for the hadronic matrix elements of high-dimensional
operators in HQE. 

HQE describes inclusive weak decays of heavy flavor hadrons. It is
a generalization of the operator product expansion (OPE) in $1/m_{Q}$,
and nonperturbative effects can be systematically studied. The starting
point of HQE is the following transition operator 
\begin{equation}
{\cal T}=i\int d^{4}x\ T[{\cal L}_{W}(x){\cal L}_{W}^{\dagger}(0)],
\end{equation}
where ${\cal L}_{W}$ is the effective weak Lagrangian governing the
decay $Q\to X_{f}$. With the help of the optical theorem the total
decay width of a hadron $H_{Q}$ containing a heavy quark $Q$ can
be given as 
\begin{equation}
\Gamma(H_{Q})=\frac{2\ {\rm Im}\langle H_{Q}|{\cal T}|H_{Q}\rangle}{2M_{H}},\label{eq:total_decay_width}
\end{equation}
where $M_{H}$ is the mass of $H_{Q}$. The right hand side of Eq.
(\ref{eq:total_decay_width}) is then calculated using OPE for the
transition operator ${\cal T}$ \cite{Cheng:2018rkz,Lenz:2014jha}
\begin{equation}
2\ {\rm Im}{\cal T}=\frac{G_{F}^{2}m_{Q}^{5}}{192\pi^{3}}\xi\left(c_{3,Q}\bar{Q}Q+\frac{c_{5,Q}}{m_{Q}^{2}}\bar{Q}g_{s}\sigma\cdot GQ+\frac{c_{6,Q}}{m_{Q}^{3}}T_{6}+\cdots\right),\label{eq:T_expansion}
\end{equation}
where $\xi$ is the relevant CKM matrix element, $T_{6}$ consists
of the four-quark operators $(\bar{Q}\Gamma q)(\bar{q}\Gamma Q)$
with $\Gamma$ representing a combination of Dirac and color matrices.

In fact, there was also a conflict between theory and experiment for
the lifetime of $\Lambda_{b}$ as early as in 1996. Taking $\tau(B^{0})=(1.519\pm0.004)$
ps in PDG2022 \cite{ParticleDataGroup:2022pth} as a benchmark, for
$\tau(\Lambda_{b})=(1.14\pm0.08)$ ps in PDG1996 \cite{Barnett:1996hr},
one can find the ratio: 
\begin{equation}
\tau(\Lambda_{b}^{0})/\tau(B^{0})=0.75\pm0.05.
\end{equation}
Theoretically, the ratio deviated from unity at the level of $20\%$
is considered to be too large. Nowadays we know that the low value
of $\tau(\Lambda_{b})/\tau(B_{d})$ or the short $\Lambda_{b}$ lifetime
was a purely experimental issue. The world averages
in PDG2022 are 
\begin{equation}
\tau(\Lambda_{b})=(1.471\pm0.009)\ {\rm ps},\qquad\tau(\Lambda_{b}^{0})/\tau(B^{0})=0.964\pm0.007,\label{eq:experiment_2022}
\end{equation}
which are in good agreement with the HQE prediction \cite{Lenz:2014jha}:
\begin{align}
\frac{\tau(\Lambda_{b})}{\tau(B_{d})}^{{\rm HQE}2014} & =1-(0.8\pm0.5)\%_{1/m_{b}^{2}}-(4.2\pm3.3)\%_{1/m_{b}^{3}}^{\Lambda_{b}}-(0.0\pm0.5)\%_{1/m_{b}^{3}}^{B_{d}}-(1.6\pm1.2)\%_{1/m_{b}^{4}}\nonumber \\
 & =0.935\pm0.054.\label{eq:HQE2014}
\end{align}

However, it can be seen from Eq. (\ref{eq:HQE2014}) that, the main
uncertainty of the lifetime ratio comes from the $1/m_{b}^{3}$ corrections
of the $\Lambda_{b}$-matrix elements. The relevant matrix elements
can be parameterized in a model-independent way \cite{Cheng:2018rkz}:
\begin{align}
\langle\Lambda_{b}|(\bar{b}q)_{V-A}(\bar{q}b)_{V-A}|\Lambda_{b}\rangle & =f_{B_{q}}^{2}m_{B_{q}}m_{\Lambda_{b}}L_{1},\nonumber \\
\langle\Lambda_{b}|(\bar{b}q)_{S-P}(\bar{q}b)_{S+P}|\Lambda_{b}\rangle & =f_{B_{q}}^{2}m_{B_{q}}m_{\Lambda_{b}}L_{2},\nonumber \\
\langle\Lambda_{b}|(\bar{b}^{\alpha}q^{\beta})_{V-A}(\bar{q}^{\beta}b^{\alpha})_{V-A}|\Lambda_{b}\rangle & =f_{B_{q}}^{2}m_{B_{q}}m_{\Lambda_{b}}L_{3},\nonumber \\
\langle\Lambda_{b}|(\bar{b}^{\alpha}q^{\beta})_{S-P}(\bar{q}^{\beta}b^{\alpha})_{S+P}|\Lambda_{b}\rangle & =f_{B_{q}}^{2}m_{B_{q}}m_{\Lambda_{b}}L_{4},\label{eq:FQO_ME}
\end{align}
where $(\bar{q}_{1}q_{2})_{V-A}\equiv\bar{q}_{1}\gamma_{\mu}(1-\gamma_{5})q_{2}$
and $(\bar{q}_{1}q_{2})_{S\pm P}\equiv\bar{q}_{1}(1\pm\gamma_{5})q_{2}$.
In the literature, a $\tilde{B}$ is usually introduced to relate
$L_{3}$ to $L_{1}$
\begin{equation}
L_{3}=-\tilde{B}\ L_{1}.\label{eq:B_tilde}
\end{equation}
The theoretical predictions for $L_{1}$, $L_{2}$ and $\tilde{B}$
are listed in Table \ref{Tab:L12_Btilde}, from which, one can see
that there are significant differences in the theoretical predictions
for $L_{1}$ and $L_{2}$. In this work, we intend to clarify this
issue. 

\begin{table}
\caption{$L_{1}$, $L_{2}$ and $\tilde{B}$ predicted by different theoretical
methods. This table is copied from Ref. \cite{Lenz:2014jha}.}
\label{Tab:L12_Btilde} %
\begin{tabular}{c|c|c|c}
\hline 
$L_{1}$  & $L_{2}$  & $\tilde{B}$  & \tabularnewline
\hline 
$-0.103(10)$  & $0.069(7)$  & $1$  & 2014 Spectroscopy update \cite{Rosner:1996fy}\tabularnewline
\hline 
$-0.22(4)$  & $0.17(2)$  & $1.21(34)$  & 1999 Exploratory Lattice \cite{DiPierro:1999tb}\tabularnewline
\hline 
$-0.22(5)$  & $0.14(3)$  & $1$  & 1999 QCDSR v1 \cite{Huang:1999xj}\tabularnewline
$-0.60(15)$  & $0.40(10)$  & $1$  & 1999 QCDSR v2 \cite{Huang:1999xj}\tabularnewline
\hline 
$-0.033(17)$  & $0.022(11)$  & $1$  & 1996 QCDSR \cite{Colangelo:1996ta}\tabularnewline
\hline 
$\approx-0.03$  & $\approx0.02$  & $1$  & 1979 Bag model \cite{Guberina:1979xw}\tabularnewline
\hline 
$\approx-0.08$  & $\approx0.06$  & $1$  & 1979 NRQM \cite{Guberina:1979xw}\tabularnewline
\hline 
\end{tabular}
\end{table}

In Ref. \cite{Shi:2019hbf}, we performed a QCD sum rules analysis
on the weak decay form factors of doubly heavy baryons to singly heavy
baryons. However, considering there were few theoretical predictions
and experimental data available to compare with, we then applied our
calculation method to study the semileptonic decay of $\Lambda_{b}\to\Lambda_{c}l\bar{\nu}$
in Ref. \cite{Zhao:2020mod}. It turns out that, our predictions for the form
factors and decay widths are consistent with those of heavy quark
effective theory (HQET) and Lattice QCD. In this work, similar computing
strategies are performed. At the QCD level, contributions from up
to dimension-6 four-quark operators are considered to obtain the
matrix elements in Eq. (\ref{eq:FQO_ME}). 

The rest of this article is arranged as follows. In Sec. II, the main
steps of QCD sum rules are presented, and in Sec. III, numerical results
are shown. For consistency, the pole residue of $\Lambda_{b}$ is also considered.
We conclude this article in the last section. 

\section{QCD sum rules analysis}

The following interpolating current is adopted for $\Lambda_{b}$:
\begin{equation}
J=\epsilon_{abc}(u_{a}^{T}C\gamma_{5}d_{b})Q_{c},
\end{equation}
where $Q$ denotes the bottom quark, $a,b,c$ are color indices, and
$C$ is the charge conjugation matrix. The correlation function is
defined as 
\begin{equation}
\Pi(p_{1},p_{2})=i^{2}\int d^{4}xd^{4}y\ e^{-ip_{1}\cdot x+ip_{2}\cdot y}\langle0|T\{J(y)\Gamma_{6}(0)\bar{J}(x)\}|0\rangle
\end{equation}
with $\Gamma_{6}$ being a four-quark operator given in Eq. (\ref{eq:FQO_ME}).
Note that, in spin space, $J(y)$, $\Gamma_{6}(0)$
and $\bar{J}(x)$ are respectively $4\times1$, $1\times1$, and $1\times4$
matrices, and thereby $\Pi(p_{1},p_{2})$ is a $4\times4$ matrix.

Following the standard procedure of QCD sum rules, one can calculate
the correlation function at the hadron level and QCD level. At the
hadron level, after inserting the complete set of baryon states, one
can obtain 
\begin{equation}
\Pi^{{\rm had}}(p_{1},p_{2})=\lambda_{H}^{2}\frac{(\slashed p_{2}+M)(a+b\gamma_{5})(\slashed p_{1}+M)}{(p_{2}^{2}-M^{2})(p_{1}^{2}-M^{2})}+\cdots,\label{eq:correlator_had}
\end{equation}
where $\lambda_{H}=\lambda_{\Lambda_{b}}$, $M=m_{\Lambda_{b}}$ are
respectively the pole residue and mass of $\Lambda_{b}$, the parameters
$a$ and $b$ are introduced to parameterize the hadronic matrix element
\begin{equation}
\langle\Lambda_{b}(q^{\prime},s^{\prime})|\Gamma_{6}|\Lambda_{b}(q,s)\rangle=\bar{u}(q^{\prime},s^{\prime})(a+b\gamma_{5})u(q,s),\label{eq:defs_a_b}
\end{equation}
and the ellipsis stands for the contribution from higher excited states.
For the forward scattering matrix element, one can show that 
\begin{align}
\langle\Lambda_{b}(q,s)|\Gamma_{6}|\Lambda_{b}(q,s)\rangle & =\bar{u}(q,s)(a+b\gamma_{5})u(q,s)\nonumber \\
 & =2\ a\ m_{\Lambda_{b}},\label{eq:matrix_element_to_a}
\end{align}
where $\bar{u}(q,s)u(q,s)=2\ m_{\Lambda_{b}}$ and $\bar{u}(q,s)\gamma_{5}u(q,s)=0$
have been used. One can see that, only the parameter $a$ in Eq. (\ref{eq:defs_a_b})
is relevant to the forward scattering matrix element. 

It can be seen from Eq. (\ref{eq:correlator_had}) that, there are
8 Dirac structures, but only (at most) 2 parameters need to be determined.
By considering the contribution of negative-parity baryons \cite{Shi:2019hbf,Zhao:2020wbw,Zhao:2020mod},
one can update Eq. (\ref{eq:correlator_had}) to 
\begin{eqnarray}
\Pi^{{\rm had}}(p_{1},p_{2}) & = & \lambda_{+}\lambda_{+}\frac{(\slashed p_{2}+M_{+})(a^{++}+b^{++}\gamma_{5})(\slashed p_{1}+M_{+})}{(p_{2}^{2}-M_{+}^{2})(p_{1}^{2}-M_{+}^{2})}\nonumber \\
 & + & \lambda_{+}\lambda_{-}\frac{(\slashed p_{2}+M_{+})(a^{+-}+b^{+-}\gamma_{5})(\slashed p_{1}-M_{-})}{(p_{2}^{2}-M_{+}^{2})(p_{1}^{2}-M_{-}^{2})}\nonumber \\
 & + & \lambda_{-}\lambda_{+}\frac{(\slashed p_{2}-M_{-})(a^{-+}+b^{-+}\gamma_{5})(\slashed p_{1}+M_{+})}{(p_{2}^{2}-M_{-}^{2})(p_{1}^{2}-M_{+}^{2})}\nonumber \\
 & + & \lambda_{-}\lambda_{-}\frac{(\slashed p_{2}-M_{-})(a^{--}+b^{--}\gamma_{5})(\slashed p_{1}-M_{-})}{(p_{2}^{2}-M_{-}^{2})(p_{1}^{2}-M_{-}^{2})}\nonumber \\
 & + & \cdots.\label{eq:correlator_had_new}
\end{eqnarray}
Here, $M_{+(-)}$ and $\lambda_{+(-)}$ respectively denote the mass
and pole residue of $\Lambda_{b}(1/2^{+(-)})$, and $a^{+-}$ is the
parameter $a$ for the positive-parity final state $\Lambda_{b}(1/2^{+})$
and the negative-parity initial state $\Lambda_{b}(1/2^{-})$, and
so forth. When arriving at Eq. (\ref{eq:correlator_had_new}) , we
have adopted the following conventions: 
\begin{align}
\langle\Lambda_{b+}(q^{\prime},s^{\prime})|\Gamma_{6}|\Lambda_{b+}(q,s)\rangle & =\bar{u}_{+}(q^{\prime},s^{\prime})(a^{++}+b^{++}\gamma_{5})u_{+}(q,s),\nonumber \\
\langle\Lambda_{b+}(q^{\prime},s^{\prime})|\Gamma_{6}|\Lambda_{b-}(q,s)\rangle & =\bar{u}_{+}(q^{\prime},s^{\prime})(a^{+-}+b^{+-}\gamma_{5})(i\gamma_{5})u_{-}(q,s),\nonumber \\
\langle\Lambda_{b-}(q^{\prime},s^{\prime})|\Gamma_{6}|\Lambda_{b+}(q,s)\rangle & =\bar{u}_{-}(q^{\prime},s^{\prime})(i\gamma_{5})(a^{-+}+b^{-+}\gamma_{5})u_{+}(q,s),\nonumber \\
\langle\Lambda_{b-}(q^{\prime},s^{\prime})|\Gamma_{6}|\Lambda_{b-}(q,s)\rangle & =\bar{u}_{-}(q^{\prime},s^{\prime})(i\gamma_{5})(a^{--}+b^{--}\gamma_{5})(i\gamma_{5})u_{-}(q,s).\label{eq:as_bs}
\end{align}
In Eq. (\ref{eq:as_bs}), these $i\gamma_{5}$ are not necessary,
but they are convenient. 

\begin{figure}[!]
\includegraphics[width=0.3\columnwidth]{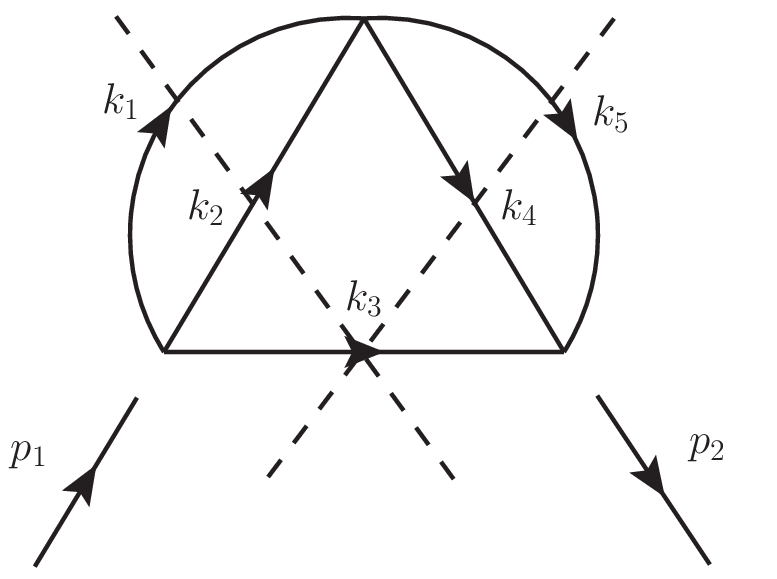}
\caption{Perturbative contribution. Cutting rules
are also shown.}
\label{fig:pert_lftm}
\end{figure}

At the QCD level, the correlation function can be written formally
as
\begin{align}
\Pi^{{\rm QCD}}(p_{1},p_{2}) & =A_{1}\ \slashed p_{2}\slashed p_{1}+A_{2}\ \slashed p_{2}+A_{3}\ \slashed p_{1}+A_{4}\nonumber \\
 & +A_{5}\ \slashed p_{2}\gamma_{5}\slashed p_{1}+A_{6}\ \slashed p_{2}\gamma_{5}+A_{7}\ \gamma_{5}\slashed p_{1}+A_{8}\ \gamma_{5}.\label{eq:correlator_QCD_formal}
\end{align}
The coefficients $A_{i}$ are then expressed as double dispersion
relations
\begin{equation}
A_{i}(p_{1}^{2},p_{2}^{2},q^{2})=\int^{\infty}ds_{1}\int^{\infty}ds_{2}\frac{\rho^{A_{i}}(s_{1},s_{2},q^{2})}{(s_{1}-p_{1}^{2})(s_{2}-p_{2}^{2})},\label{eq:correlator_QCD}
\end{equation}
where the spectral function $\rho^{A_{i}}(s_{1},s_{2},q^{2})$ can
be calculated using Cutkosky cutting rules, see Fig. \ref{fig:pert_lftm}.
Sum rules are obtained by equating Eq. (\ref{eq:correlator_had_new})
to Eq. (\ref{eq:correlator_QCD_formal}) and then using quark-hadron
duality to eliminate the contribution of excited states. Furthermore,
by equating the coefficients of the same Dirac structure, one can
have 8 equations to solve 8 unknown parameters $a^{\pm\pm}$ and $b^{\pm\pm}$.
Especially, after performing the Borel transform, one can arrive at
\begin{equation}
a^{++}=\frac{\{M_{-}^{2},M_{-},M_{-},1\}.\{{\cal B}A_{1},{\cal B}A_{2},{\cal B}A_{3},{\cal B}A_{4}\}}{\lambda_{+}^{2}(M_{+}+M_{-})^{2}}\exp\left(\frac{2M_{+}^{2}}{T^{2}}\right),\label{eq:a_plus_plus}
\end{equation}
where ${\cal B}A_{i}$ are doubly Borel transformed coefficients
\begin{equation}
{\cal B}A_{i}=\int^{s_{0}}ds_{1}\int^{s_{0}}ds_{2}\ \rho^{A_{i}}(s_{1},s_{2},q^{2})\exp\left(-\frac{s_{1}+s_{2}}{T^{2}}\right),
\end{equation}
 and $s_{0}$ and $T^{2}$ are respectively continuum threshold parameter
and Borel parameter. 

\begin{figure}[!]
\includegraphics[width=1.0\columnwidth]{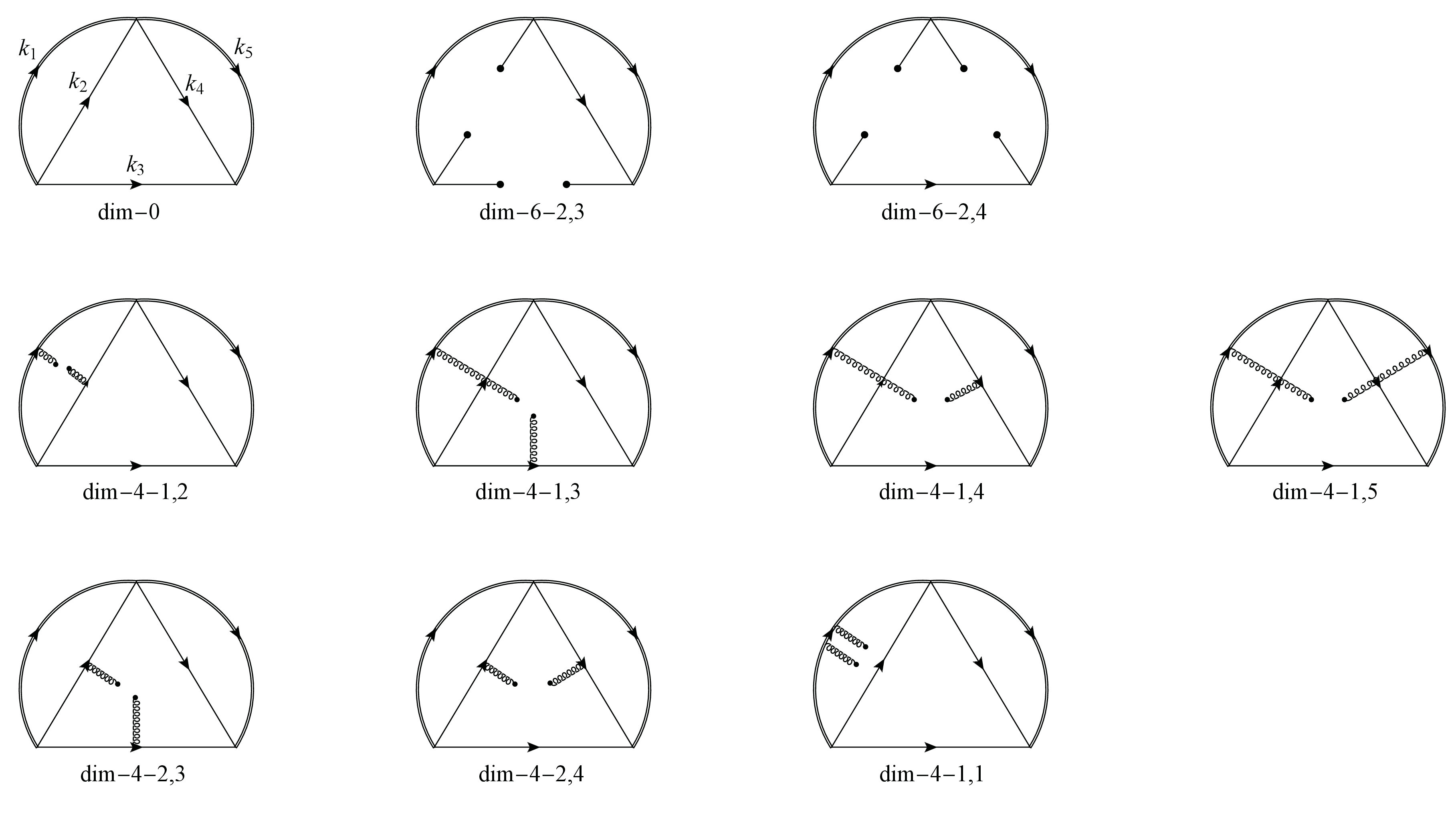}
\caption{All nonzero,
``independent'' diagrams considered in this work. }
\label{fig:OPE_diagrams}
\end{figure}

In this work, contributions from up to dimension-6 four-quark operators are considered
at the QCD level. \footnote{Readers should not confuse the four-quark operators in HQE
with those at the QCD level in QCD sum rules.}
For the matrix elements in Eq. (\ref{eq:FQO_ME}),
we find that contributions from quark condensate (dimension-3) and
quark-gluon condensate (dimension-5) are proportional to the mass
of $u/d$ quark, which is taken to be zero in this work. All nonzero,
``independent'' diagrams are shown in Fig. \ref{fig:OPE_diagrams}.
Here, ``independent'' is explained as follows. 
For example, diagram dim-4-2,5
with quark 2 and quark 5 each emitting a gluon
is equal to diagram dim-4-1,4,
therefore the former is not listed in Fig. \ref{fig:OPE_diagrams}. 

\subsection{The pole residue}

\begin{figure}[!]
\includegraphics[width=0.6\columnwidth]{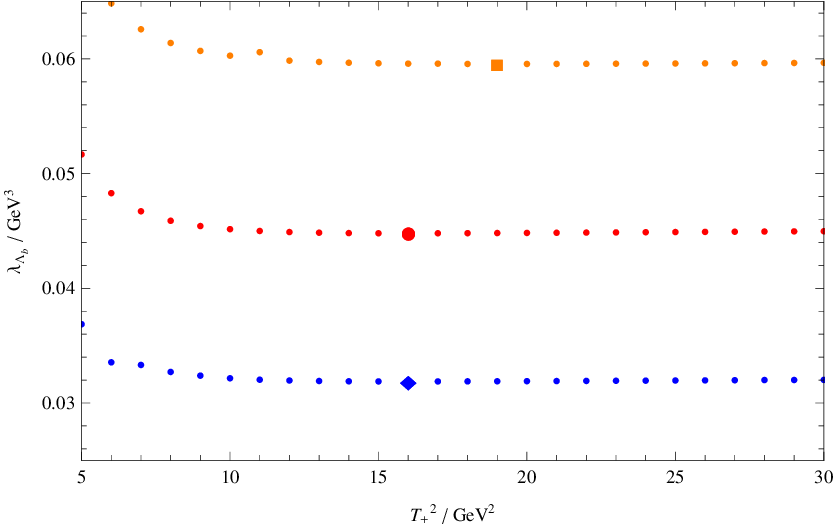}
\caption{
The pole residue $\lambda_{\Lambda_{b}}$ as a function of the Borel parameter $T_{+}^{2}$.
The red, orange, and blue dots respectively correspond to the renormalization scale $\mu=m_{b}$, $\mu=6\ {\rm GeV}$,
and $\mu=3\ {\rm GeV}$. The minimum points are marked, and they correspond to the experimental mass of 
$\Lambda_{b}$ via Eq. (\ref{eq:mass_formula}). 
}
\label{fig:pole_residue}
\end{figure}

As can be seen in Eq. (\ref{eq:a_plus_plus}), the pole residue of
$\Lambda_{b}$ is an indispensable input. For consistency, in this
work, we also perform an analysis on the pole residue of $\Lambda_{b}$,
whose sum rule is \cite{Zhao:2020mod}
\begin{equation}
(M_{+}+M_{-})\lambda_{+}^{2}\exp(-M_{+}^{2}/T_{+}^{2})=\int^{s_{0}}ds\ (M_{-}\rho^{A}+\rho^{B})\exp(-s/T_{+}^{2}),\label{eq:pole_residue_sum_rule}
\end{equation}
from which, one can obtain the mass formula for $\Lambda_{b}$
\begin{equation}
M_{+}^{2}=\frac{\int^{s_{0}}ds\ (M_{-}\rho^{A}+\rho^{B})\ s\ \exp(-s/T_{+}^{2})}{\int^{s_{0}}ds\ (M_{-}\rho^{A}+\rho^{B})\exp(-s/T_{+}^{2})}.\label{eq:mass_formula}
\end{equation}
Eq. (\ref{eq:mass_formula}) can be viewed as a constraint to Eq.
(\ref{eq:pole_residue_sum_rule}). 
Following Ref. \cite{Jamin:2001fw},
in this work, we use Eq. (\ref{eq:mass_formula}) to determine the
continuum threshold parameter $s_{0}$.
Specifically, for a set of fixed parameters (renormalization scale, condensate parameters, etc.),
the optimal $(s_{0}, T_{+}^{2})$ are obtained through the following procedure:
\begin{enumerate}
\item For a trial $s_{0}$, plot the pole residue curve with respect to the
Borel parameter $T_{+}^{2}$ using Eq. (\ref{eq:pole_residue_sum_rule}).
Find the minimum point $T_{+}^{2}$ on the curve. (See Fig. \ref{fig:pole_residue}
for some intuitive impressions.)
\item Substitute the set of ($s_{0}$, $T_{+}^{2}$) obtained in step 1
into Eq. (\ref{eq:mass_formula}) to calculate the baryon mass, and compare it with the
experimental value. If equal (within a small error range), terminate;
otherwise, go to step 1.
\end{enumerate}
For different input parameters, one can obtain different optimal ($s_{0}$,
$T_{+}^{2}$), which are listed in Table \ref{Tab:results}. 

In this work, contributions from up to dimension-6 four-quark operators
are also considered for the sum rule in Eq. (\ref{eq:pole_residue_sum_rule}). 

\section{Numerical results}

\subsection{Inputs}

Our main inputs in numerical calculation can be found in Table \ref{Tab:results}.
We take the $\overline{{\rm MS}}$ mass for the bottom quark \cite{ParticleDataGroup:2022pth}
and neglect the mass of $u/d$ quark. The condensate parameters are
taken from Ref. \cite{Colangelo:2000dp}. The renormalization scale
is taken as $\mu_{b}=3\sim6\ {\rm {\rm GeV}}$ with $m_{b}(m_{b})$
the central value \cite{Jamin:2001fw}, from which, one can estimate the dependence of
the calculation results on the renormalization scale. 

\subsection{The pole residue and the continuum threshold parameter}

Our predictions for the pole residue $\lambda_{\Lambda_{b}}$, together with the continuum
threshold parameter $s_{0}$ can be found in Table \ref{Tab:results}.
A similar investigation has also been performed in our previous work
\cite{Zhao:2020mod}, however, in this work, more contributions from
higher dimensional operators are considered. Numerically, the predictions
in this work are close to those in Ref. \cite{Zhao:2020mod}.
This is essentially because the contributions from higher dimensional
operators are small. A comprehensive study of the pole residues of
antitriplet heavy baryons can be found in Ref. \cite{Wang:2010fq}. 

\begin{table}
\caption{Our predictions of the pole residue $\lambda_{\Lambda_{b}}$, $L_{1,2,3,4}$
defined in Eq. (\ref{eq:FQO_ME}), and $\tilde{B}\equiv-L_{3}/L_{1}$.
For a set of fixed parameters (renormalization scale, condensate
parameters, etc.), the continuum threshold parameter $s_{0}$ can
be determined, and then the quantities we are interested in can also
be determined -- by requiring them to have as little dependence on
the Borel parameter as possible. }
\label{Tab:results} %
\begin{tabular}{c|c|c|c|c|c|c}
\hline 
 & Central value & $\begin{array}{c}
m_{b}/{\rm GeV}=\\
4.18\pm0.03
\end{array}$ & $\begin{array}{c}
\mu/{\rm GeV}=6.0\\
\mu/{\rm GeV}=3.0
\end{array}$ & $\begin{array}{c}
\langle\bar{q}q\rangle(1\,{\rm GeV})/{\rm GeV}^{3}\\
=-(0.24\pm0.01)^{3}
\end{array}$ & $\begin{array}{c}
\langle g_{s}^{2}GG\rangle/{\rm GeV}^{4}=\\
0.47\times(1.0\pm0.3)
\end{array}$ & $\begin{array}{c}
m_{\Lambda_{b}}/{\rm GeV}=\\
5.620\pm0.001
\end{array}$\tabularnewline
\hline 
\hline 
$s_{0}/{\rm GeV}^{2}$ & $36.08$ & $\begin{array}{c}
36.08\\
36.01
\end{array}$ & $\begin{array}{c}
36.09\\
36.03
\end{array}$ & $\begin{array}{c}
36.09\\
36.04
\end{array}$ & $\begin{array}{c}
36.08\\
36.08
\end{array}$ & $\begin{array}{c}
36.09\\
36.07
\end{array}$\tabularnewline
\hline 
$T_{+}^{2}/{\rm GeV}^{2}$ & $16$ & $\begin{array}{c}
16\\
17
\end{array}$ & $\begin{array}{c}
19\\
16
\end{array}$ & $\begin{array}{c}
19\\
14
\end{array}$ & $\begin{array}{c}
16\\
16
\end{array}$ & $\begin{array}{c}
16\\
16
\end{array}$\tabularnewline
\hline 
$\lambda_{\Lambda_{b}}$ & $0.0448$ & $\begin{array}{c}
0.0429\\
0.0465
\end{array}$ & $\begin{array}{c}
0.0596\\
0.0319
\end{array}$ & $\begin{array}{c}
0.0452\\
0.0443
\end{array}$ & $\begin{array}{c}
0.0448\\
0.0448
\end{array}$ & $\begin{array}{c}
0.0449\\
0.0447
\end{array}$\tabularnewline
\hline 
$\Delta\lambda_{\Lambda_{b}}$ & - - & $\begin{array}{c}
-0.0019\\
+0.0017
\end{array}$ & $\begin{array}{c}
+0.0148\\
-0.0129
\end{array}$ & $\begin{array}{c}
+0.0004\\
-0.0005
\end{array}$ & $\begin{array}{c}
-0.0000\\
+0.0000
\end{array}$ & $\begin{array}{c}
+0.0001\\
-0.0001
\end{array}$\tabularnewline
\hline 
\hline 
$T^{2}/{\rm GeV}^{2}$ & $100$ & $\begin{array}{c}
150\\
180
\end{array}$ & $\begin{array}{c}
180\\
180
\end{array}$ & $\begin{array}{c}
150\\
120
\end{array}$ & $\begin{array}{c}
100\\
90
\end{array}$ & $\begin{array}{c}
150\\
120
\end{array}$\tabularnewline
\hline 
$L_{1}$ & $-0.143$ & $\begin{array}{c}
-0.139\\
-0.144
\end{array}$ & $\begin{array}{c}
-0.171\\
-0.119
\end{array}$ & $\begin{array}{c}
-0.150\\
-0.136
\end{array}$ & $\begin{array}{c}
-0.144\\
-0.143
\end{array}$ & $\begin{array}{c}
-0.143\\
-0.143
\end{array}$\tabularnewline
\hline 
$\Delta L_{1}$ & - - & $\begin{array}{c}
+0.004\\
-0.001
\end{array}$ & $\begin{array}{c}
-0.027\\
+0.024
\end{array}$ & $\begin{array}{c}
-0.007\\
+0.007
\end{array}$ & $\begin{array}{c}
-0.001\\
+0.001
\end{array}$ & $\begin{array}{c}
-0.000\\
+0.000
\end{array}$\tabularnewline
\hline 
\hline 
$T^{2}/{\rm GeV}^{2}$ & $100$ & $\begin{array}{c}
210\\
240
\end{array}$ & $\begin{array}{c}
260\\
220
\end{array}$ & $\begin{array}{c}
270\\
120
\end{array}$ & $\begin{array}{c}
100\\
90
\end{array}$ & $\begin{array}{c}
150\\
120
\end{array}$\tabularnewline
\hline 
$L_{3}$ & $0.151$ & $\begin{array}{c}
0.147\\
0.151
\end{array}$ & $\begin{array}{c}
0.177\\
0.128
\end{array}$ & $\begin{array}{c}
0.156\\
0.144
\end{array}$ & $\begin{array}{c}
0.154\\
0.148
\end{array}$ & $\begin{array}{c}
0.151\\
0.151
\end{array}$\tabularnewline
\hline 
$\Delta L_{3}$ & - - & $\begin{array}{c}
-0.004\\
-0.000
\end{array}$ & $\begin{array}{c}
+0.026\\
-0.024
\end{array}$ & $\begin{array}{c}
+0.005\\
-0.007
\end{array}$ & $\begin{array}{c}
+0.003\\
-0.003
\end{array}$ & $\begin{array}{c}
-0.000\\
-0.000
\end{array}$\tabularnewline
\hline 
\hline 
$\begin{array}{c}
\tilde{B}=\\
-L_{3}/L_{1}
\end{array}$ & $1.057$ & $\begin{array}{c}
1.056\\
1.049
\end{array}$ & $\begin{array}{c}
1.041\\
1.073
\end{array}$ & $\begin{array}{c}
1.040\\
1.060
\end{array}$ & $\begin{array}{c}
1.073\\
1.040
\end{array}$ & $\begin{array}{c}
1.056\\
1.056
\end{array}$\tabularnewline
\hline 
$\Delta\tilde{B}$ & - - & $\begin{array}{c}
-0.000\\
-0.007
\end{array}$ & $\begin{array}{c}
-0.016\\
+0.017
\end{array}$ & $\begin{array}{c}
-0.016\\
+0.003
\end{array}$ & $\begin{array}{c}
+0.017\\
-0.017
\end{array}$ & $\begin{array}{c}
-0.001\\
-0.000
\end{array}$\tabularnewline
\hline 
\hline 
$T^{2}/{\rm GeV}^{2}$ & $160$ & $\begin{array}{c}
170\\
200
\end{array}$ & $\begin{array}{c}
190\\
140
\end{array}$ & $\begin{array}{c}
220\\
190
\end{array}$ & $\begin{array}{c}
160\\
160
\end{array}$ & $\begin{array}{c}
140\\
180
\end{array}$\tabularnewline
\hline 
$L_{2}$ & $0.0746$ & $\begin{array}{c}
0.0728\\
0.0750
\end{array}$ & $\begin{array}{c}
0.0882\\
0.0632
\end{array}$ & $\begin{array}{c}
0.0780\\
0.0713
\end{array}$ & $\begin{array}{c}
0.0758\\
0.0733
\end{array}$ & $\begin{array}{c}
0.0745\\
0.0744
\end{array}$\tabularnewline
\hline 
$\Delta L_{2}$ & - - & $\begin{array}{c}
-0.0018\\
+0.0004
\end{array}$ & $\begin{array}{c}
+0.0136\\
-0.0114
\end{array}$ & $\begin{array}{c}
+0.0033\\
-0.0033
\end{array}$ & $\begin{array}{c}
+0.0011\\
-0.0013
\end{array}$ & $\begin{array}{c}
-0.0001\\
-0.0002
\end{array}$\tabularnewline
\hline 
\hline 
$T^{2}/{\rm GeV}^{2}$ & $160$ & $\begin{array}{c}
170\\
200
\end{array}$ & $\begin{array}{c}
190\\
140
\end{array}$ & $\begin{array}{c}
220\\
190
\end{array}$ & $\begin{array}{c}
160\\
160
\end{array}$ & $\begin{array}{c}
140\\
180
\end{array}$\tabularnewline
\hline 
$L_{4}$ & $-0.0764$ & $\begin{array}{c}
-0.0746\\
-0.0767
\end{array}$ & $\begin{array}{c}
-0.0896\\
-0.0654
\end{array}$ & $\begin{array}{c}
-0.0797\\
-0.0732
\end{array}$ & $\begin{array}{c}
-0.0781\\
-0.0745
\end{array}$ & $\begin{array}{c}
-0.0763\\
-0.0762
\end{array}$\tabularnewline
\hline 
$\Delta L_{4}$ & - - & $\begin{array}{c}
+0.0018\\
-0.0003
\end{array}$ & $\begin{array}{c}
-0.0133\\
+0.0110
\end{array}$ & $\begin{array}{c}
-0.0033\\
+0.0032
\end{array}$ & $\begin{array}{c}
-0.0017\\
+0.0018
\end{array}$ & $\begin{array}{c}
+0.0001\\
+0.0002
\end{array}$\tabularnewline
\hline 
\end{tabular}
\end{table}

\subsection{The four-quark operator matrix elements}

\begin{figure}[htbp]
	\centering
	\begin{minipage}{0.75\linewidth}
		\centering
		\includegraphics[width=1.0\linewidth]{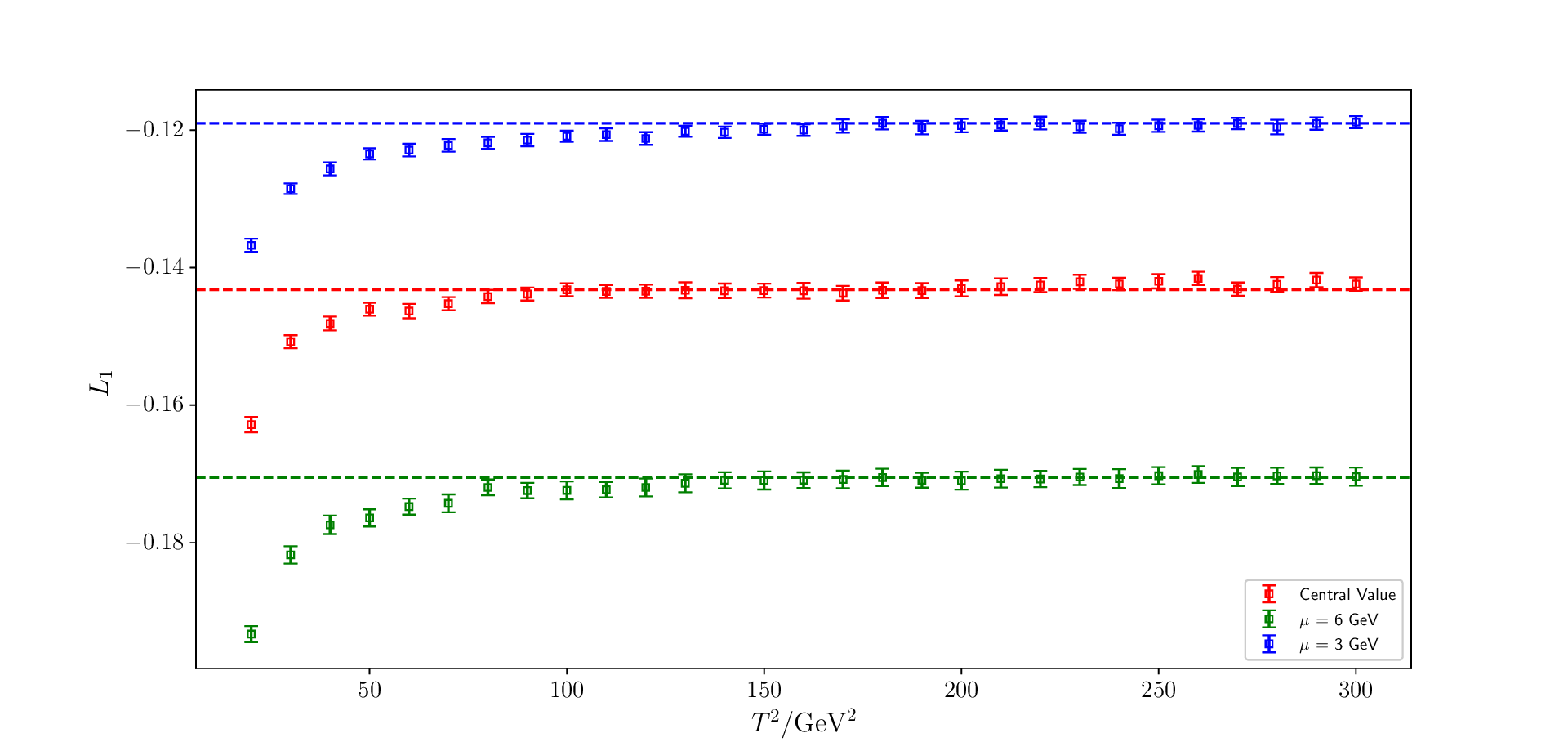}
		\includegraphics[width=1.0\linewidth]{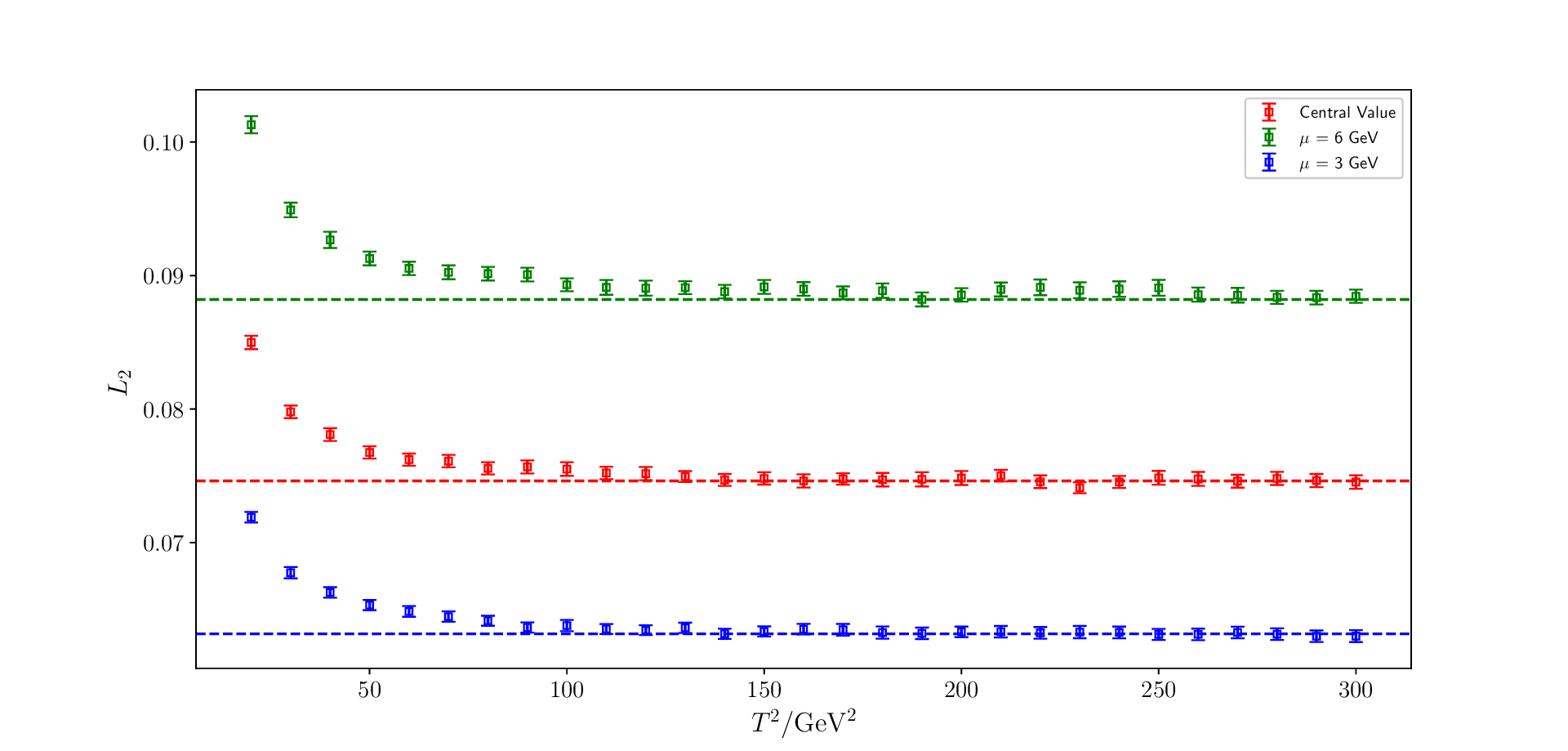}
		\includegraphics[width=1.0\linewidth]{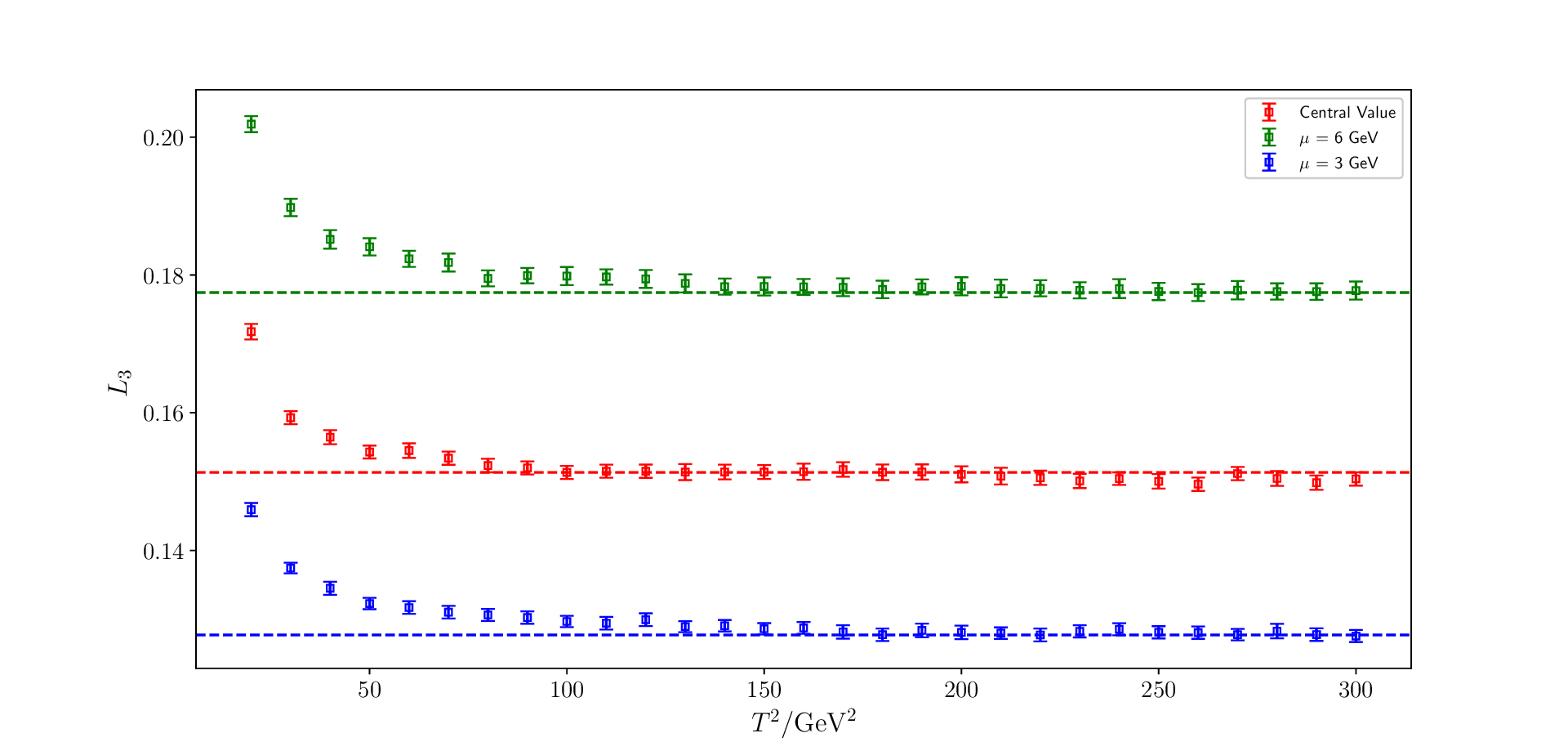}
		\includegraphics[width=1.0\linewidth]{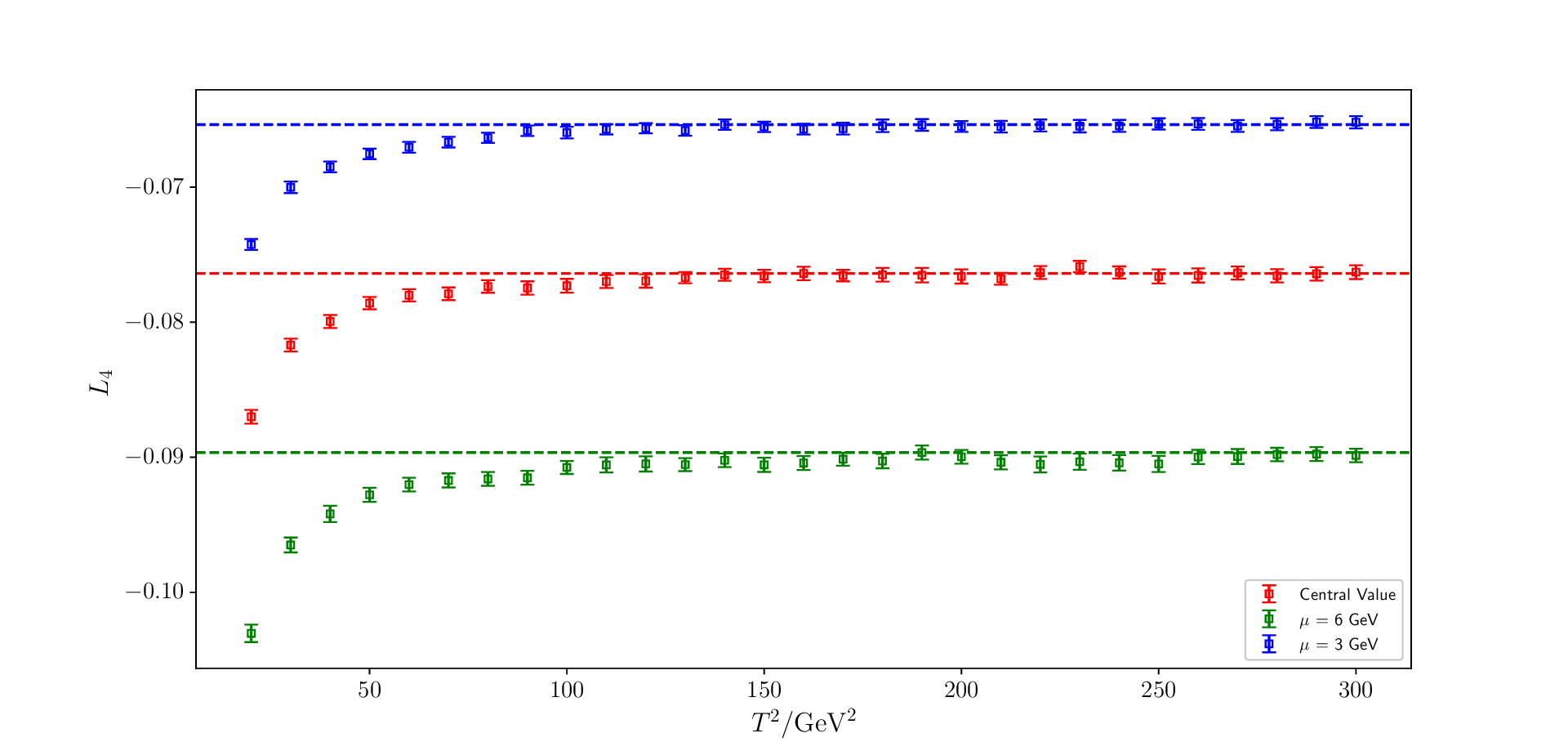}
	\end{minipage}
	\caption{$L_{1,2,3,4}$ as functions of the Borel parameter $T^{2}$.}
	\label{fig:L1234}
\end{figure}

The four-quark operator matrix elements in Eq. (\ref{eq:FQO_ME})
are all proportional to some parameter $L_{i}$. In Fig. \ref{fig:L1234}, we plot the curves of $L_{1,2,3,4}$ as functions
of the Borel parameter $T^{2}$, on which, stability region can be
found. In these figures, we have also evaluated the dependence of $L_{1,2,3,4}$
on the renormalization scale. The corresponding results
are summarized in Table \ref{Tab:results}.

Adding all the uncertainties from various input parameters in quadrature,
our final results of $L_{1,2,3,4}$ and $\tilde{B}\equiv-L_{3}/L_{1}$
are respectively
\begin{align}
L_{1} & =-0.143\pm0.028,\nonumber \\
L_{2} & =+0.0746\pm0.0142,\nonumber \\
L_{3} & =+0.151\pm0.027,\nonumber \\
L_{4} & =-0.0764\pm0.0139,
\end{align}
and 
\begin{equation}
\tilde{B}=1.057\pm0.030,
\end{equation}
where $m_{B_{q}}=5.280\ {\rm GeV}$ and $f_{B_{q}}=186\ {\rm MeV}$
\cite{Cheng:2018rkz} have been used.
It can be seen from Table \ref{Tab:L12_Btilde} that, our results
are close to those of the spectroscopy update of Ref. \cite{Rosner:1996fy} in 2014.

Some comments are in order. 
\begin{itemize}
\item The spectral density in Eq. (\ref{eq:correlator_QCD}) also depends
on $q^{2}\equiv(p_{1}-p_{2})^{2}$. For the forward scattering
matrix elements of interest, $q^{2}$ is taken to be 0. 
\item In Eq. (\ref{eq:FQO_ME}), the only difference between the first matrix
element $\langle\Lambda_{b}|(\bar{b}q)_{V-A}(\bar{q}b)_{V-A}|\Lambda_{b}\rangle$
and the third matrix element $\langle\Lambda_{b}|(\bar{b}^{\alpha}q^{\beta})_{V-A}(\bar{q}^{\beta}b^{\alpha})_{V-A}|\Lambda_{b}\rangle$
is in the color space. For all the diagrams in Fig. \ref{fig:OPE_diagrams} except for
diagrams (dim-4-1,4), (dim-4-1,5), and (dim-4-2,4), the color factors
for the two matrix elements differ only by a negative sign. Considering
that the contributions of these three gluon condensate diagrams are
small, $\tilde{B}\equiv-L_{3}/L_{1}$ only deviates slightly by 1. 
\item One can see from Table \ref{Tab:results}
that, the uncertainty caused by $\mu$ dependence is dominant,
which is about $20\%$. As a comparison, in Ref. \cite{Jamin:2001fw},
the authors performed a QCD sum rules analysis on the decay constants
of $B$ and $B_{s}$ mesons. $\mu$ is also taken in the range $3\sim6$ GeV
with $m_{b}(m_{b})$ as the central value. For the perturbative spectral function,
contributions from up to $\alpha_{s}^{2}$ order are considered. Scale
dependence is also the main source of error, which is approximately
$10\%$. It can be expected that, when higher-order corrections
are considered, the scale dependence should become smaller. 
\item In Fig. \ref{fig:L1234}, we have also shown the
calculation errors, which are essentially small. Nevertheless, when
we try to find the stability region on the curve, these small errors also play a role. 
\item As can be seen in Table \ref{Tab:results} that,
for a set of fixed parameters (quark mass, renormalization scale, condensate
parameters), the continuum threshold parameter $s_{0}$ can
be determined, and then the quantities we are interested in can also
be determined -- by requiring them to have as little dependence on
the Borel parameter as possible.
\item In our opinion, the continuum threshold parameter $s_{0}$ is the
most important parameter in QCD sum rules. In fact, once $s_{0}$
is fixed, the quantity that we are interested in is almost determined
by searching for stability region. In the literature, there exist
at least two approaches to determine $s_{0}$. One approach is to
empirically select $\sqrt{s_{0}}$ approximately 0.5 GeV larger than
the ground state mass. Another approach is to determine $s_{0}$ through
the mass formula. We have adopted the latter approach in this work. The basic
logic behind doing so is that the mass formula can be seen as a constraint
to the sum rule of two-point correlation function. Of course, this
comes at the cost of abandoning the predictive power of hadron mass.
\end{itemize}

\section{Conclusions}

Heavy quark expansion can nicely explain the lifetime of $\Lambda_{b}$.
However, there still exist sizable uncertainties in the four-quark
operator matrix elements of $\Lambda_{b}$ in $1/m_{b}^{3}$ corrections,
which describe the spectator effects. In this work, these four-quark operator matrix
elements are investigated using full QCD
sum rules. At the QCD level, contributions from up to dimension-6
four-quark operators are considered. Stable Borel region can be found.
We have also considered the uncertainties from various input parameters,
and find that the main source of error is scale dependence,
which is about 20\%. Our results are close to those of the spectroscopy
update of Ref. \cite{Rosner:1996fy} in 2014. Our method
of calculating high-dimensional operator matrix elements
is promising to be used to resolve the $\Omega_{c}$ lifetime puzzle. 

\section*{Acknowledgements}

The authors are grateful to Profs. Pietro Colangelo, Yue-Long Shen,
Wei Wang, Zhi-Gang Wang, Fu-Sheng Yu, and Dr. Yu-Ji Shi for valuable
discussions. This work is supported in part by scientific research start-up fund
for Junma program of Inner Mongolia University, scientific research
start-up fund for talent introduction in Inner Mongolia Autonomous
Region, and National Natural Science Foundation of China under Grant
No. 12065020.

\end{document}